\documentclass[twocolumn,conference,letterpaper]{IEEEtran}
\usepackage[left=0.75in,right=0.75in,top=1in,bottom=0.75in]{geometry}

\usepackage{cite}
\usepackage{amssymb,amsmath,latexsym,amsfonts,amsthm,mathtools}

\usepackage{graphicx}
\usepackage{float,subcaption}
\usepackage{textcomp}
\usepackage{xcolor}
\definecolor{darkpastelpurple}{rgb}{0.59, 0.44, 0.84}
\usepackage{hyperref}
\hypersetup{colorlinks, breaklinks, citecolor=blue, linkcolor=blue, urlcolor=blue}
\usepackage[nameinlink]{cleveref}
\Crefformat{figure}{#2Fig.~#1#3}
\Crefmultiformat{figure}{Figs.~#2#1#3}{ and~#2#1#3}{, #2#1#3}{ and~#2#1#3}

\usepackage{tikz}
\usetikzlibrary{automata, shapes, arrows, calc, arrows.meta, fit, positioning}

\theoremstyle{plain}
\newtheorem{theorem}{Theorem}

\newtheorem{proposition}[theorem]{Proposition}
\newtheorem*{problem*}{Problem}
\theoremstyle{remark}
\newtheorem{remark}{Remark}

\theoremstyle{definition}

\usepackage{mathrsfs}

\usepackage{newtxmath}
\usepackage{booktabs}
\usepackage{duckuments}
\usepackage{bm}
\usepackage{xspace}

\newcommand{\TIGER}{{\sc Et-Mapg}\xspace}
\newcommand{\ATTIGER}{{\sc Aet-Mapg}\xspace}

\IEEEoverridecommandlockouts

\begin{document}
	\title{Adaptive Event-Triggered Policy Gradient for Multi-Agent Reinforcement Learning}
	\author{Umer Siddique, Abhinav Sinha,~\IEEEmembership{Senior Member,~IEEE}, and Yongcan Cao,~\IEEEmembership{Senior Member,~IEEE}
		\thanks{U. Siddique and Y. Cao are with the Unmanned Systems Lab, Department of Electrical and Computer Engineering, The University of Texas at San Antonio, San Antonio, TX 78249, USA. (e-mails: muhammadumer.siddique@my.utsa.edu, yongcan.cao@utsa.edu). A. Sinha is with the GALACxIS Lab, Department of Aerospace Engineering and Engineering Mechanics, University of Cincinnati, OH 45221, USA. (email: abhinav.sinha@uc.edu).}
	}

	\maketitle
	\thispagestyle{empty}
	
\begin{abstract}
    Conventional multi-agent reinforcement learning (MARL) methods rely on time-triggered execution, where agents sample and communicate actions at fixed intervals. This approach is often computationally expensive and communication-intensive. To address this limitation, we propose \TIGER (Event-Triggered Multi-Agent Policy Gradient reinforcement learning), a framework that jointly learns an agent's control policy and its event-triggering policy. Unlike prior work that decouples these mechanisms, \TIGER integrates them into a unified learning process, enabling agents to learn not only what action to take but also when to execute it. For scenarios with inter-agent communication, we introduce \ATTIGER, an attention-based variant that leverages a self-attention mechanism to learn selective communication patterns. \ATTIGER empowers agents to determine not only when to trigger an action but also with whom to communicate and what information to exchange, thereby optimizing coordination. Both methods can be integrated with any policy gradient MARL algorithm. Extensive experiments across diverse MARL benchmarks demonstrate that our approaches achieve performance comparable to state-of-the-art, time-triggered baselines while significantly reducing both computational load and communication overhead.
\end{abstract}

\begin{IEEEkeywords}
    Multi-agent reinforcement learning, event-triggered learning and control, self-attention, data-driven control.
\end{IEEEkeywords}
	
\section{Introduction}\label{sec:introduction}
Event-triggered control (ETC)~\cite{10.1201/b19013} is an approach in which signals are updated or exchanged only when a certain state or output condition is met, rather than at fixed, periodic intervals. The main goal of ETC is to reduce communication and computation while maintaining closed-loop performance, in contrast to time-triggered control (TTC). ETC has been widely studied~\cite{selivanov2015event,9838043,9849174,https://doi.org/10.1002/rnc.6024}, where most of these methods assume access to an accurate system dynamics or model. Although this may be possible in small-scale simulated environments, it's often unrealistic or nearly impossible in complex real-world applications. 

To mitigate this model dependence, data-driven ETC methods have been recently proposed. These methods include discrete-time formulations~\cite{10042042,https://doi.org/10.1002/rnc.6024,8619335, 10147364,9838043} but continuous-time approaches are scarce, e.g., \cite{9849174}. For linear time-invariant systems, several works simplify learning by ignoring disturbances during offline data collection~\cite{10042042,https://doi.org/10.1002/rnc.6024,9838043}. In contrast, the work in \cite{10147364} proposes a more realistic method that includes designing the controller and trigger from a single batch of noisy data, thereby accounting for disturbances and measurement errors. Moreover, the work in \cite{9849174} incorporates disturbances both during learning and in closed-loop operation via a dynamic triggering strategy that guarantees $\mathcal{L}_2$ stability.

Reinforcement learning (RL) has achieved strong empirical results in sequential decision-making and control, including robotics~\cite{bucsoniu2018reinforcement,bertsekas2019reinforcement,ibarz2021train,siddique2024deep}. Yet, most of the RL works focus on designing a time-triggered control policy, often overlooking communication cost. While a few model-free RL-based ETC methods exist~\cite{VAMVOUDAKIS2018412,6889787,8619335,8183439,siddique2025adaptive}, they are developed for a single agent. In practice, many systems are inherently multi-agent systems with tight bandwidth constraints. In MARL, multiple agents are interacting, learning, and coordinating with each other to solve a shared task, making event-triggered learning even more important, where multiple agents should act and communicate only when necessary.

Communication in MARL is essential for coordination and efficient problem-solving, especially under partial observability. Early work in MARL introduced deep distributed recurrent Q-networks (DDRQN), e.g., ~\cite{foerster2016learning}, which demonstrate that agents can learn communication protocols for coordination. Building on this, the work in \cite{foerster2016learningb} proposed Reinforced Inter-Agent Learning (RIAL) and Differentiable Inter-Agent Learning (DIAL) to learn communication end to end, and the authors in \cite{kim2019learning} investigated communication scheduling with relational inductive biases. However, these approaches often rely on specialized communication networks or extensive parameter sharing, which can be costly at scale. Inspired by the success of self-attention~\cite{bahdanau2014neural, vaswani2017attention}, we instead learn compact, attention-based messages where agents, during the learning phase, compute attention scores over shared representations and exchange only the relevant information.

ETC has also been explored in MARL to some extent. ETCNet~\cite{hu2021event} reduces bandwidth by sending messages only when necessary, but all agents still interact with the environment at every time step, and triggering applies only to inter-agent communication. ETMAPPO~\cite{feng2023approximating} integrates ETC with multi-agent proximal policy optimization algorithms via a Beta strategy to compress transmitted information and accelerate convergence in specific UAV environments. Although their model-free multi-agent PPO method performs well in the anti-UAV jamming scenario, their approach also applies ETC only to communication among agents.

To address these limitations, we propose Event-Triggered Multi-Agent Policy Gradient reinforcement learning (\TIGER), which jointly learns both the control action head and an event-trigger head for each agent and decides when to update the action and/or communicate. Unlike approaches that learn triggering conditions and control actions with separate policies~\cite{6889787,lu2023event,chen2022reinforcement}, \TIGER employs a single shared network with two heads. Due to this, \TIGER reduces the policy network parameters, latency, and improves efficiency, which is crucial when scaling many agents in MARL. When inter-agent communication is allowed, we further introduce \ATTIGER, an attention-based variant of \TIGER that leverages self-attention to facilitate selective, learned message passing during training. As a consequence of the proposed approach, the communication graph in \ATTIGER is inherently sparse since an agent resamples an action or transmits messages only when its triggering condition is satisfied. Otherwise, it reuses its previous action and suppresses messaging. This design reduces communication and computation while supporting stable and efficient deployment. Our main contributions are summarized as follows: 
\begin{itemize}
    \item We propose \TIGER, a method that jointly learns a control action head and an event-trigger head for each agent. The trigger head determines when a new action should be sampled, thereby providing an improvement over approaches that learn triggering and control with separate policies.
    \item We further propose \ATTIGER, an attention-based variant of \TIGER that uses self-attention as a communication mechanism during training, which can improve the coordination efficiency.
    \item We demonstrate the generality of \TIGER and \ATTIGER by integrating them with three state-of-the-art multi-agent policy gradient algorithms, including IPPO~\cite{de2020independent}, MAPPO~\cite{yu2022surprising}, and IA2C~\cite{papoudakis2021benchmarking}.
    \item Through our extensive experiments, we show that \TIGER and \ATTIGER match the performance of standard MARL algorithms while reducing computation cost by up to 50\%.
\end{itemize}

\section{Background and Preliminaries} \label{sec:preliminaries}
We consider a multi-agent system of $N$ agents, indexed by $i \in \mathcal{I} = \{1, \ldots, N\}$, communicating over a fully connected undirected graph $\mathcal{G} = (\mathcal{I}, \mathcal{E})$. The dynamics of each agent are governed by a discrete-time nonlinear equation given by
\begin{equation}
    \mathbf{x}_{i,k+1} = \mathbf{f}_i(\mathbf{x}_{i,k}, \mathbf{u}_{i,k}, \{\mathbf{x}_{j,k}\}_{j \in \mathcal{N}_i}),
\end{equation}
where $\mathbf{x}_{i,k}$ is the state of agent $i$, $\mathbf{u}_{i,k}$ is its control input, and $\mathcal{N}_i$ is the set of its neighbors. We model the problem of multi-agent event-triggered learning as a decentralized partially observable Markov decision process (Dec-POMDP) which is defined by a tuple $\langle \mathcal{I}, \mathcal{X}, \{\mathcal{U}_i\}, \mathcal{P}, r, \{\mathcal{O}_i\}, \mathcal{Z}, \gamma\rangle$, where $\mathcal{I} = \{1, \ldots, N\}$ is the set of $N$ agents. The true state of the environment is $\mathbf{x} \in \mathcal{X}$. At each timestep $k$, every agent $i \in \mathcal{I}$ selects an action $\mathbf{u}_{i,k} \in \mathcal{U}_i$ from its individual action set. This forms a joint action $\mathbf{u}_k=(\mathbf{u}_{1,k}, \ldots, \mathbf{u}_{N,k}) \in \boldsymbol{\mathcal{U}}$, where $\boldsymbol{\mathcal{U}} = \times_{i \in \mathcal{I}} \mathcal{U}_i$ is the joint action space. The joint action governs the state transition according to the probability function $\mathcal{P}(\mathbf{x}_{k+1} \mid \mathbf{x}_k, \mathbf{u}_k)$. In this cooperative setting, the team of agents receives a single shared reward, $\mathbf{r}_k = \mathbf{r}(\mathbf{x}_k, \mathbf{u}_k)$. The agents, however, do not observe the true state $\mathbf{x}_k$. Instead, after the transition to state $\mathbf{x}_{k+1}$, each agent $i$ receives a private observation $\mathbf{o}_{i,k+1} \in \mathcal{O}_i$. The joint observation $\mathbf{o}_{k+1} = (\mathbf{o}_{1,k+1}, \ldots, \mathbf{o}_{N,k+1})$ is determined by the observation function $\mathcal{Z}(\mathbf{o}_{k+1} \mid \mathbf{x}_{k+1}, \mathbf{u}_k)$. Finally, $\gamma \in [0, 1)$ is the discount factor. 

Each agent maintains a local action-observation history, $\tau_{i,k} = (\mathbf{o}_{i,0}, \mathbf{u}_{i,0}, \ldots, \mathbf{u}_{i,k-1}, \mathbf{o}_{i,k})$. Actions are chosen according to a local, stochastic policy, $\mathbf{u}_{i,k} \sim \pi_i(\cdot \mid \tau_{i,k})$. The team of agents aims to learn a joint policy $\boldsymbol{\pi}$ that factorizes into the local policies
\begin{equation}
    \boldsymbol{\pi}(\mathbf{u}_k \mid \boldsymbol{\tau}_k) = \prod_{i=1}^N \pi_i(\mathbf{u}_{i,k} \mid \tau_{i,k}),
\end{equation}
where $\boldsymbol{\tau}_k = (\tau_{1,k}, \ldots, \tau_{N,k})$ is the joint history. The objective is to find a joint policy that maximizes the expected discounted return
\begin{equation}
    \bm J(\bm \pi) = \mathbb{E}_{\bm \pi, \mathcal{P}, \mathcal{Z}} \left[\sum_{k=0}^\infty \gamma^k \mathbf{r}(\mathbf{x}_k, \mathbf{u}_k)\right].
\end{equation}
\begin{remark}
    In this model, we adopt the centralized training with decentralized execution (CTDE) paradigm (see \cite{foerster2016learning,foerster2016learningb,de2020independent,yu2022surprising,papoudakis2021benchmarking}), which combines the benefits of having global information during training and decentralized scalability at execution. In CTDE, agents are assumed to have access to the full state of the system during training, which helps in mitigating non-stationarity in dynamic environments (a common challenge that arises when multiple agents interact with a shared environment, causing the dynamics to shift from the perspective of each agent). However, during execution, agents' actions only depend on their local observations, which is crucial for real-world deployment where global state information is usually unavailable or may not be possible to get due to bandwidth, latency, or privacy constraints.
\end{remark}

To conserve computational and communication resources, we depart from the standard time-triggered paradigm. Instead, we employ an event-triggered scheme where each agent decides independently when to compute and broadcast a new action. Each agent $i$ maintains its own sequence of event times $\{t^i_j\}_{j \in \mathbb{N}}$. At an event time $t^i_j$, it updates its history $\tau_{i, t^i_j}$ and computes a new action by sampling from its policy, $\mathbf{u}_{i, t^i_j} \sim \pi_i(\cdot \mid \tau_{i, t^i_j})$. This action is then held constant until the next event
\begin{equation}
    \mathbf{u}_{i,k} = \mathbf{u}_{i, t^i_j}\,, \quad \forall k \in [t^i_j, t^i_{j+1}).
\end{equation}
The next event time, $t^i_{j+1}$, is determined by a local triggering rule based on an error signal, $\mathbf{e}_{i,k} = \mathbf{x}_{i,k} - \mathbf{x}_{i, t^i_j}$, such that
\begin{equation}
    t^i_{j+1} = \inf \{ k > t^i_j \mid \mathcal{T}_i(\mathbf{x}_{i,k}, \mathbf{e}_{i,k}) \geq 0 \}.
\end{equation}
The goal is to co-design the policies $\{\pi_i\}$ and triggering functions $\{\mathcal{T}_i\}$ to maximize the expected return $J(\boldsymbol{\pi})$ while significantly reducing the frequency of policy evaluations and network communication.

Previously, the approaches often relied on parameter sharing (e.g., latent information) or specialized architectures for message exchange, both of which typically assume access to full state information or sufficient bandwidth (see e.g.,~\cite{foerster2016learning,foerster2016learningb} and references therein). To address these limitations, we leverage \emph{attention mechanisms} as a communication tool. Self-attention computes queries $\mathcal{Q}$, keys $\mathcal{K}$, and values $\mathcal{V}$, and evaluates attention
\begin{align}
    \mathscr{A}(\mathcal{Q}, \mathcal{K}, \mathcal{V}) = \text{softmax} \left( \frac{\mathcal{Q} \mathcal{K}^\top}{\sqrt{d_k}} \right) \mathcal{V},
\end{align}
where $d_k$ is the dimensionality of $\mathcal{K}$. Multi-head self-attention extends this by projecting $\mathcal{Q}, \mathcal{K}, \mathcal{V}$ into multiple subspaces, applying attention in parallel, and concatenating the outputs. This allows agents to selectively focus on relevant information and capture diverse interaction patterns. By integrating this multi-head attention as a communication mechanism with event-triggered learning, we let MARL agents decide \emph{what}, \emph{how}, and \emph{when} to communicate. Note that we assume that the agents communicate over a complete undirected graph at the triggering instants, which indicates a bidirectional messaging sharing at the triggering instants only.

\begin{figure*}[t]
\centering
    \begin{subfigure}[t]{0.32\textwidth}
    \centering
    \includegraphics[width=\linewidth]{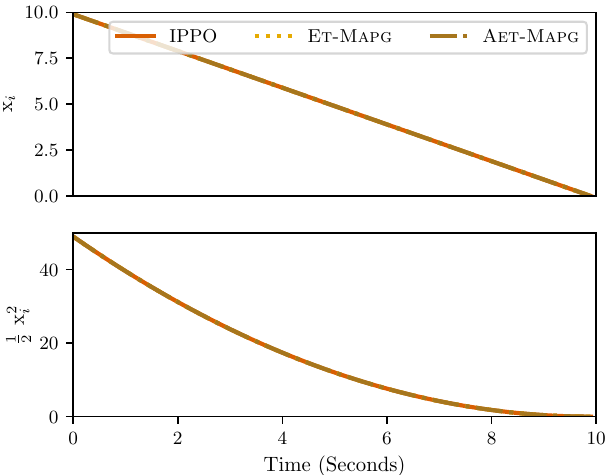}
    \caption{State and Lyapunov.}
    \label{fig:sintlyap}
    \end{subfigure}
    \begin{subfigure}[t]{0.32\textwidth}
    \centering
    \includegraphics[width=\linewidth]{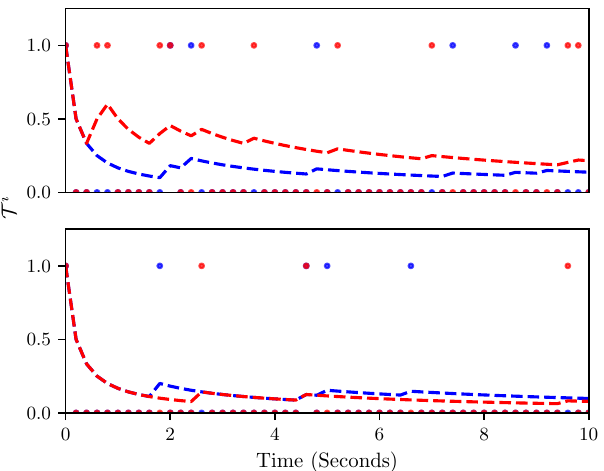}
    \caption{Triggering instants and moving avg.}
    \label{fig:sintcomm}
    \end{subfigure}
    \begin{subfigure}[t]{0.32\textwidth}
    \centering
    \includegraphics[width=\linewidth]{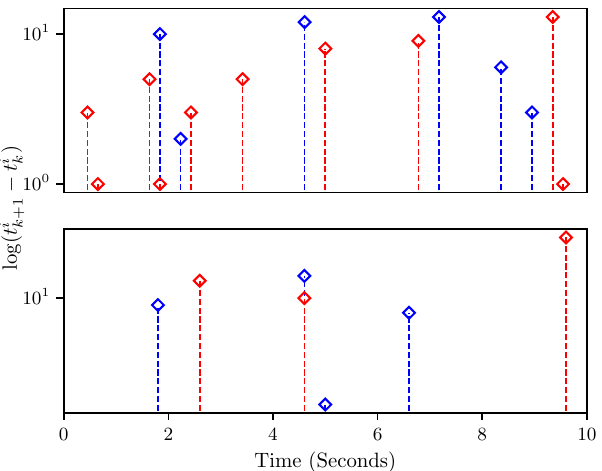}
    \caption{Inter-event time.}
    \label{fig:sintzeno}
    \end{subfigure}
\caption{Performance comparison between IPPO and our proposed methods in the perturbed multi-agent single integrators. Subfigures (b) and (c) show \TIGER (top) and \ATTIGER (bottom) results, respectively.} 
\label{fig:sint}
\end{figure*}

\section{Proposed Approach}
We now present the proposed event-triggered framework in multi-agent settings. Since our objective is to address large-scale multi-agent problems with high-dimensional (or possibly even continuous state-action spaces in general), we focus on multi-agent deep policy gradient methods such as IPPO~\cite{de2020independent}, MAPPO~\cite{yu2022surprising}, and IA2C~\cite{papoudakis2021benchmarking}. Among these methods, IPPO and IA2C learn independent actors and critics for each agent, whereas MAPPO employs a centralized critic with decentralized actors. Since all these methods share the same actor-critic architecture, the proposed framework can be easily extended to any of them. 

In a cooperative setting, the goal of agents is to learn a joint policy $\bm{\pi}_{\bm{\theta}} = (\pi_{1,\theta},\dots,\pi_{n,\theta})$ that maximizes the discounted sum of rewards given by 
\begin{align}\label{eq:problem}
\max_{\bm \pi_{\bm \theta}} \bm J(\bm \pi_{\bm \theta}) = \max_{\bm \pi_{\bm \theta}} \left( 
\mathbb{E}_{\bm \pi, \mathcal{P}, \mathcal{Z}} \left[ \sum_{k=0}^\infty \gamma^k \mathbf{r}(\mathbf{x}_k, \mathbf{u}_k) \right] \right),
\end{align}
where $\mathcal{P}$ denotes the environment transition dynamics.
\begin{remark}
    In independent learning settings, where agents are learning independently by observing their local observations and performing actions individually, this decomposes into maximizing each agent’s expected return in the form of
\begin{align}
J_i(\pi_{i,\theta}) = \mathbb{E}_{\pi_{i,\theta}} \Big[ \sum_{k=0}^\infty \gamma_i^k r_{i,k} \Big], \quad \forall~ i=1,\dots,n.
\end{align}
\end{remark}

To maximize \eqref{eq:problem}, standard MARL algorithms update policies at every time step, which is equivalent to TTC. Traditional ETC designs have attempted to subdue the limitations of time-triggered execution, although they design triggering conditions manually (e.g., based on state deviations) and treat them separately from the control policy \cite{6889787,8619335,lu2023event}. To address this issue, we propose \TIGER, an independent learning MARL algorithm in which each agent $i$ jointly learns both its control action $\mathbf{u}_{i,k}$ and its triggering condition $\mathcal{T}_i$ such that
\begin{align}
\left(\mathcal{T}_i,\mathbf{u}_{i,k}\right) = \pi_{i,\theta}\left(\mathbf{z}_{i,k}, \tau_{i,k}\right),~\forall \mathbf{z}_{i,k}\in\mathcal{Z}.\label{eq:parout}
\end{align}
\begin{proposition}
    Let each agent $i$ in a multi-agent system, at a given timestep $k$, maintain a local observation $\mathbf{z}_{i,k}$ and a state-action history $\tau_{i,k}$. Consider a parametric policy $\pi_{i,\theta}$ that jointly outputs the control action $\mathbf{u}_{i,k}$ and the triggering decision $\mathcal{T}_i$ as given in \eqref{eq:parout}. If agent $i$ maximizes the expected cumulative reward with a triggering regularization
    \begin{align}
J_i(\pi_{i,\theta}) 
= \mathbb{E}_{\pi_{i,\theta}} \left[\sum_{k=0}^\infty \gamma_i^k \, r_i(\mathbf{z}_{i,k}, \mathbf{u}_{i,k}) 
\right] - \Psi \cdot \mathbb{I}(\mathcal{T}_i = 1),
\end{align}
where $\Psi$ penalizes frequent triggering, and $\mathbb{I}$ is the indicator function, then the optimal policy $\pi_{i,\theta}^\ast = \arg\max_{\pi_{i,\theta}} J_i(\pi_{i,\theta})$ simultaneously learns what action to take and when to update.
\end{proposition}
\begin{remark}
  This unified approach reduces the model complexity by avoiding the need for hand-crafted triggers and improves sample efficiency by allowing agents to dynamically decide both {what action to take} and {when to update}. In fact, if $\mathcal{T}_i = 1$ at every timestep, then a higher penalty discourages frequent policy updates and incentivizes efficient communication and computation by trading off performance with triggering frequency.   
\end{remark}
Following the policy gradient theorem~\cite{sutton1999policy}, the gradient for each agent $i$ is computed as
\begin{align}
\nabla_{i,\theta} J(\pi_{i,\theta}) =&  \mathbb{E}_{\pi_{i,\theta}} \left[ \mathbf{A}_{i,k}(\mathbf{z}_{i,k}, \mathbf{u}_{i,k}) \nabla_{i,\theta} \log \pi_{i,\theta}(\mathbf{u}_{i,k} \mid \mathbf{z}_{i,k}) \right] \nonumber \\&
- \Psi \cdot \mathbb{I}(\mathcal{T}_i = 1),
\end{align}
where $\mathbf{A}_{i,k}$ is the vector advantage function operated componentwise at a given timestep. Positive components indicate improvement for that particular agent, and negative components indicate a worse policy than the baseline.
\begin{remark}
    Our proposed method \TIGER is model-agnostic and is sufficiently general to be extended to a large class of MARL algorithms.
\end{remark}
As an illustrative example, we first extend IPPO \cite{de2020independent} to incorporate the proposed framework by employing the clipped surrogate PPO objective augmented with the triggering penalty
\begin{align}
\mathbb{E}_{\pi_{i,\theta}}\Big[ \min\left(\rho_{i,\theta} \mathbf{A}_{i,k}^\text{IPPO}, \bar{\rho}_{i,\theta} \mathbf{A}_{i,k}^\text{IPPO}\right) \Big] 
- \Psi \cdot \mathbb{I}(\mathcal{T}_i = 1),
\end{align}
where $\rho_{i,\theta}$ and $\bar{\rho}_{i,\theta}$ denote PPO’s ratio and clipping terms for agent $i$. The advantage $\mathbf{A}_{i,k}^\text{IPPO}$ is estimated using TD($\lambda$) given by
\begin{align*}
    \mathbf{A}_{i,k}^\text{IPPO} =& \sum_{k} (\gamma_i \lambda)^{k-1} \left(r_i(\mathbf{z}_{i,k}, \mathbf{u}_{i,k})\right. \\ &\left.+ \gamma \mathbf{V}_{i,\theta}(\mathbf{z}_{i,k+1}) - \mathbf{V}_{i,\theta}(\mathbf{z}_{i,k}) \right),
\end{align*}
where $\mathbf{V}_{i,\theta}$ is the vector value function for agent $i$. As a consequence of this modular design, \TIGER can be extended to other policy gradient MARL algorithms.

For instance, with MAPPO~\cite{yu2022surprising}, the centralized critic enables advantage estimation as \begin{align*}
    \mathbf{A}_{i,k}^\text{MAPPO} =& \sum_{k} (\gamma_i \lambda)^{k-1} \left(r_i(\mathbf{z}_{i,k}, \mathbf{u}_{i,k}) \right.\nonumber\\
    &\left.+ \gamma \mathbf{V}_{i,\theta}(\mathbf{x}_{k+1}) - \mathbf{V}_{i,\theta}(\mathbf{x}_{k}) \right),
\end{align*} 
where $\mathbf{x}_k$ denotes the global state. Similarly, in IA2C, the advantage reduces to $$\mathbf{A}_{i,k}^\text{IA2C} = r_i(\mathbf{z}_{i,k}, \mathbf{u}_{i,k} - \mathbf{V}_{i,\theta}(\mathbf{z}_{i,k}).$$ Therefore, our approach provides a general event-triggered extension to any policy gradient MARL algorithms.

While \TIGER efficiently learn both the control action and policy and triggering condition, it assumes agents act independently without explicit communication. However, many cooperative tasks require coordination, where agents can communicate to reach a consensus or mitigate non-stationarity~\cite{papoudakis2019dealing}. To this end, we propose \ATTIGER, a variant of \TIGER which integrates event-triggered communication with self-attention.
\begin{proposition}
    Let \TIGER be an independent learning framework in which each agent $i$ jointly learns its control action $\mathbf{u}_{i,k}$ and the triggering decision $\mathcal{T}_i$. \ATTIGER is a variant of \TIGER that integrates event-triggered communication with self-attention, such that when $\mathcal{T}_i=1$, agent $i$ broadcasts its learned message to all other agents over $\mathcal{G}$. Each agent aggregates received messages via multi-head self-attention
    \begin{align}
    b_i = \sum_{j} \alpha_{ij} v_j; \quad 
    \alpha_{ij} = \text{softmax}\left(\frac{[\mathcal{Q}]_i [\mathcal{K}_j]^\top}{\sqrt{d_k}}\right),
\end{align}
where $\alpha_{ij}$ denotes the attention weight of agent $i$ attending to agent $j$. This mechanism allows agents to selectively exchange information only when triggering conditions are met, ensuring both coordination and efficiency.
\end{proposition}
The aggregated message $b_i$ is then fused into the policy network.  With multiple attention heads ($h=4$ in our experiments), agents capture diverse interaction patterns, improving robustness in cooperative settings.

\begin{figure*}[t]
\centering
    \begin{subfigure}[t]{0.32\textwidth}
    \centering
    \includegraphics[width=\linewidth]{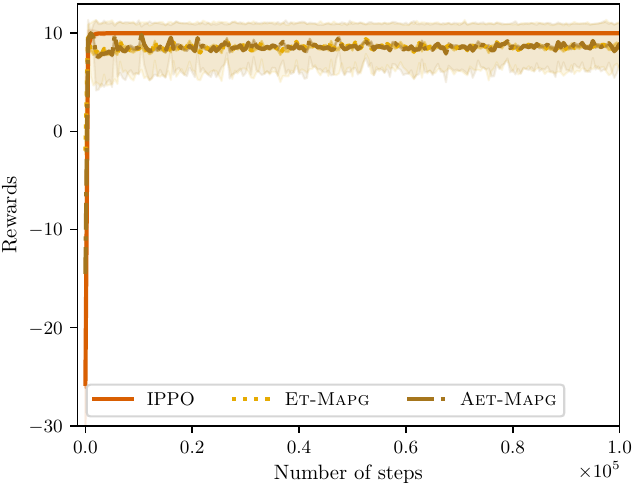}
    \caption{Rewards.}
    \label{fig:matrixrewards}
    \end{subfigure}
    \begin{subfigure}[t]{0.32\textwidth}
    \centering
    \includegraphics[width=\linewidth]{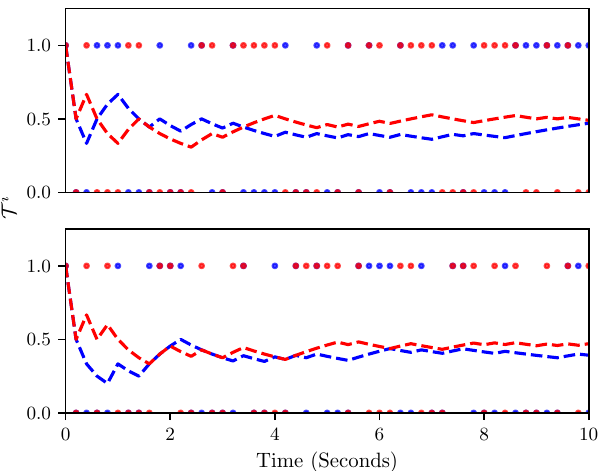}
    \caption{Triggering instants and moving avg.}
    \label{fig:matrixcomm}
    \end{subfigure}
    \begin{subfigure}[t]{0.32\textwidth}
    \centering
    \includegraphics[width=\linewidth]{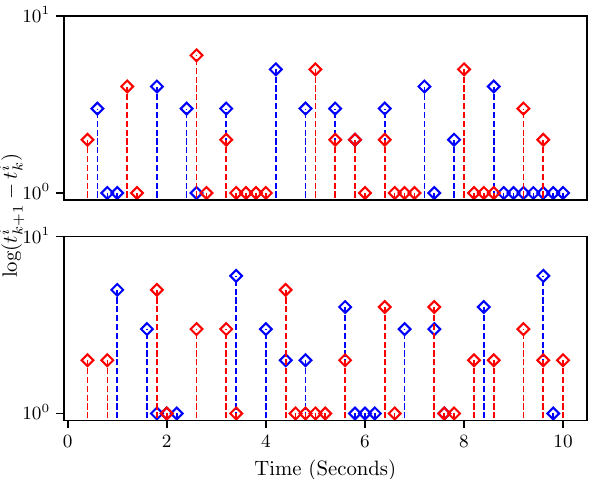}
    \caption{Inter event time.}
    \label{fig:matrixzeno}
    \end{subfigure}
\caption{Performance comparison between IPPO and our proposed methods in the repeated matrix game. Subfigures (b) and (c) show TIGER (top) and ATT-TIGER (bottom) results, respectively.} 
\label{fig:matrix}
\end{figure*}

\section{Experimental Results} \label{sec:results}
To demonstrate the effectiveness of our proposed methods, we evaluate them across diverse environments with different levels of complexity and communication demands. Specifically, we consider (i) perturbed chain of single integrators, (ii) the repeated penalty matrix game~\cite{claus1998dynamics}, and (iii) multi-agent particle environments (MPE)~\cite{mordatch2017emergence}, which are a collection of 2D simulated physics environments that require cooperation or competition among agents. Via these experiments, we evaluate both control and coordination abilities of our methods.  For all experiments, we perform hyperparameter optimization and report results for the best-performing configurations. Furthermore, for the sake of reproducibility, we run each experiment with five different seeds and report the mean performance. Unless otherwise specified, we benchmark \TIGER and \ATTIGER against IPPO~\cite{de2020independent} in the main experiments and show the generality of our methods with other policy gradient methods in ablation studies.

In the first case, the agents seek stabilization of the origin from different initial conditions. At each time step $k$, agent $i$ updates according to $\mathbf{x}_{i,k+1} = \mathbf{x}_{i,k} + \mathbf{u}_{i,k}T_s$,
where $T_s$ is the sampling time. The reward function for agent $i$ is given by the improvement in state convergence, which means encouraging an agent to reduce the absolute value of the state (i.e., getting closer to the equilibrium point) while penalizing large or unnecessary control actions. In a cooperative setting, the goal of all agents is to reach the equilibrium point.

The results in \Cref{fig:sint} show that both proposed methods successfully stabilize the system while requiring substantially fewer action updates. The initial state value, which is $10$, is driven to equilibrium in each case. However, \TIGER and \ATTIGER achieve this while reducing triggering events by at least 60\%. \Cref{fig:sintlyap} (top) illustrates the state trajectories of the system, while the bottom plot shows the decay of the standard quadratic Lyapunov function over time for a representative agent $i$. In both cases, \TIGER and \ATTIGER match IPPO performance .~\Cref {fig:sintcomm} demonstrates the communication frequency with the policy, where the triggering frequencies for \TIGER (top) and \ATTIGER (bottom) are significantly lower than those of the IPPO baseline, which triggers policy interaction at each time step. Here, blue and red triggering instants and moving average curves correspond to the two agents in the environment. Finally, \Cref{fig:sintzeno} shows the inter-event times across agents, demonstrating that they remain strictly positive and adaptive, avoiding Zeno behavior while ensuring stable convergence. 

\begin{figure*}[t]
\centering
    \begin{subfigure}[t]{0.32\textwidth}
    \centering
    \includegraphics[width=\linewidth]{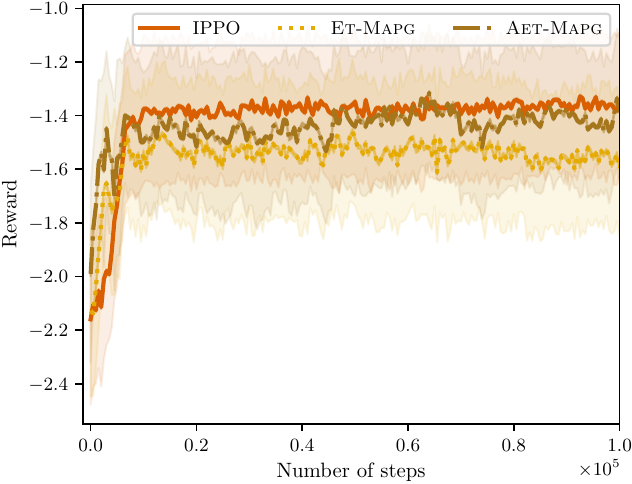}
    \caption{Rewards.}
    \label{fig:refrewards}
    \end{subfigure}
    \begin{subfigure}[t]{0.32\textwidth}
    \centering
    \includegraphics[width=\linewidth]{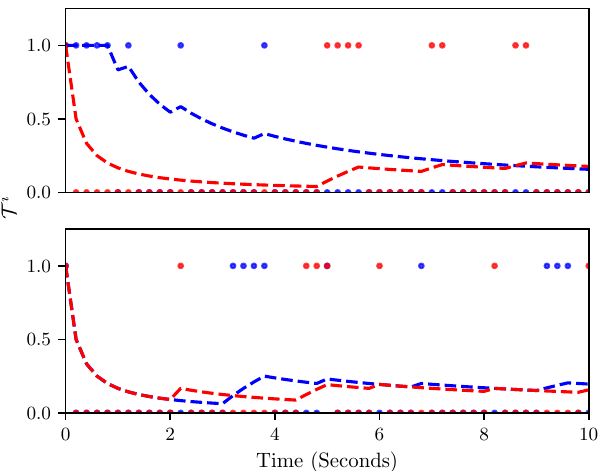}
    \caption{Triggering instants and moving avg.}
    \label{fig:refcomm}
    \end{subfigure}
    \begin{subfigure}[t]{0.32\textwidth}
    \centering
    \includegraphics[width=\linewidth]{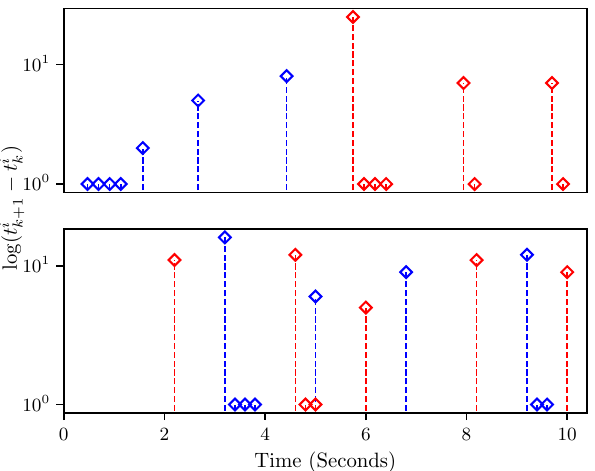}
    \caption{Inter event time.}
    \label{fig:refzeno}
    \end{subfigure}
\caption{Performance comparison between IPPO and our proposed methods in the Simple Reference MPE. Subfigures (b) and (c) show TIGER (top) and ATT-TIGER (bottom) results, respectively.} 
\label{fig:ref}
\end{figure*}

\begin{figure*}[t]
\centering
    \begin{subfigure}[t]{0.32\textwidth}
    \centering
    \includegraphics[width=\linewidth]{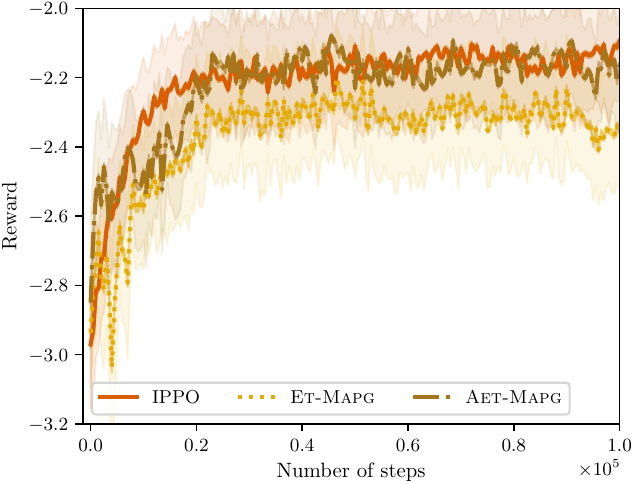}
    \caption{Rewards.}
    \label{fig:mperewards}
    \end{subfigure}
    \begin{subfigure}[t]{0.32\textwidth}
    \centering
    \includegraphics[width=\linewidth]{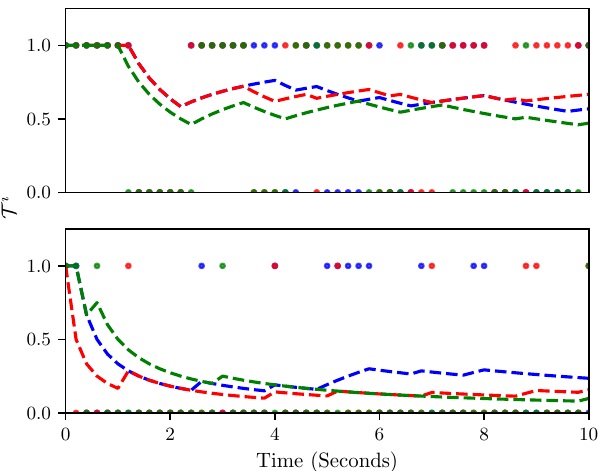}
    \caption{Triggering instants and moving avg.}
    \label{fig:mpecomm}
    \end{subfigure}
    \begin{subfigure}[t]{0.32\textwidth}
    \centering
    \includegraphics[width=\linewidth]{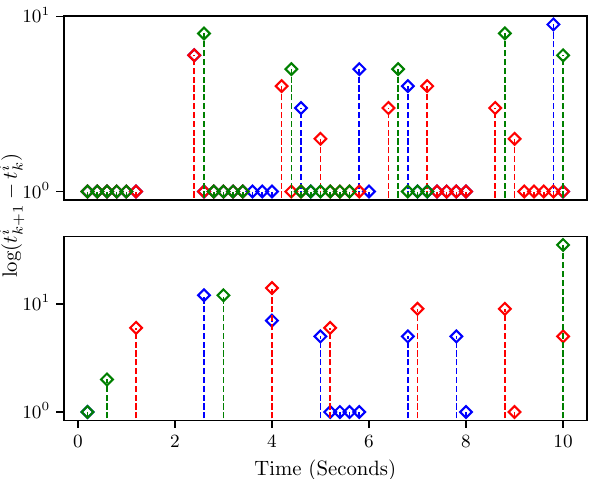}
    \caption{Inter event time.}
    \label{fig:mpezeno}
    \end{subfigure}
\caption{Performance comparison between IPPO and our proposed methods in the Simple Spread MPE. Subfigures (b) and (c) show TIGER (top) and ATT-TIGER (bottom) results, respectively.} 
\label{fig:spread}
\end{figure*}

We now test the proposed algorithms on the repeated penalty matrix game by Claus and Boutilier~\cite{claus1998dynamics}, which is a two-agent cooperative setting defined by the payoff matrix
\begin{align*}
\begin{bmatrix} 
\ell & 0 & 10 \\ 
0 & 2 & 0 \\ 
10 & 0 & \ell 
\end{bmatrix}, 
\end{align*}
with $\ell \leq 0$ (set to $\ell=-100$ for more complexity). At each timestep, agents observe their local state, perform their own actions, and receive rewards. Specifically, agent 1 selects a row as its action and agent 2 selects a column, producing a joint payoff from the corresponding matrix entry (the entry where their choices intersect). To earn the highest reward of 10, the agents must select the correct combination of actions simultaneously. Alternatively, certain actions are considered safe, which guarantee a smaller but reliable reward of 2, regardless of what the other agent chooses. At the same time, failure to cooperate may incur the penalty of $\ell=-100$. Each episode length is $25$, where each position in the matrix remains the same and agents only rely on a constant observation, which makes the environment stateless aside from episode progression. As the penalty is strongly negative, even small deviations from cooperative behavior lead to catastrophic consequences, and agents fall into the local Nash equilibria (i.e., agents keep choosing the safe actions that yield rewards equal to 2).

\Cref{fig:matrixrewards} demonstrates the rewards achieved by our proposed methods compared to the IPPO, indicating that our methods perform comparably to the IPPO. Although our methods have slightly lower rewards, their triggering frequency is significantly lower than the standard IPPO, which triggers at every time step (see~\Cref{fig:matrixcomm}). The inter-event time in~\Cref{fig:matrixzeno} confirms that our methods maintain strictly positive intervals, adapting the triggering schedule dynamically and avoiding Zeno behavior. These results show that both \TIGER and \ATTIGER preserve performance while reducing communication overhead, even in sparse environments with high-risk coordination.

Our third evaluation domain is the Multi-Agent Particle Environments (MPE), a widely used suite of continuous 2D tasks where agents with simple dynamics must solve cooperative or competitive problems under partial observability. Since our focus is on cooperative MARL, we consider two tasks: \emph{Simple Reference} and \emph{Simple Spread}. These tasks require agents to balance control efficiency with inter-agent communication and make them suitable for testing event-triggered methods. In these environments, each agent observes its local position and velocity, relative positions of landmarks and other agents, and optional communication inputs. The action space of each agent includes \emph{no-action, move-left, move-right, move-down, and move-up}. In \emph{Simple Reference}, the environment consists of two agents and three landmarks. Each landmark is a fixed circular location, and each agent is assigned a private target landmark known only to the other agent. In this environment, agents act both as speakers and listeners, and their goal is to navigate to their assigned targets. The reward function in this environment consists of local and global rewards, where a local reward for each agent is the negative distance to its own target, and the global reward is defined as the average distance of all agents to their respective targets. In \emph{Simple Spread}, there are three agents and three landmarks. The goal in this environment is to cover all landmarks while avoiding collisions. Again, the reward function consists of both local and global components, where a local reward is a penalty of $-1$ for each collision and a global reward is the negative sum of the minimum distances from each landmark to the closest agent, encouraging agents to spread out and efficiently cover all landmarks. 

\Cref{fig:refrewards,fig:mperewards} present the learning curves in both environments and show that while \TIGER significantly reduces communication frequency, it may converge to slightly lower rewards compared to the baseline. This is because, in an event-triggered setting, and especially in environments where communication is necessary, the proposed methods allow an efficient communication mechanism. Instead of constant updates, agents learn to share information only at the most critical moments. This targeted communication helps them build a better model of the environment's dynamics and achieve tighter coordination as a team.  This is evident as \ATTIGER leverages selective communication to match the reward performance of IPPO, while also achieving the most resource-efficient behavior. Communication frequencies shown in \Cref{fig:refcomm,fig:mpecomm} demonstrate that both methods substantially reduce communication frequencies, where \ATTIGER performs the best. ~\Cref{fig:refzeno,fig:mpezeno} present the inter-event time. Once again, these results show that triggering intervals remain strictly positive and dynamically adjusted over time, which avoids Zeno behavior. These results demonstrate that our proposed event-triggered methods also generalize to high-dimensional, partially observable MPE domains, achieving both performance and resource-efficiency.

\begin{figure}[h]
    \centering
    \includegraphics[width=0.95\linewidth]{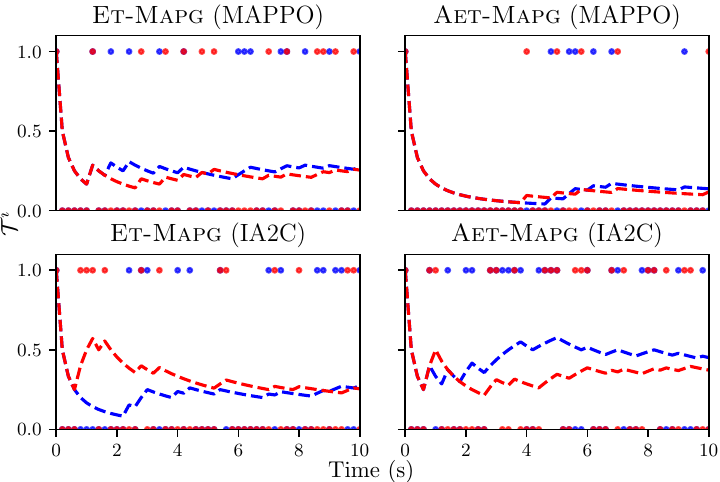}
    \caption{Communication frequencies of event-triggered MAPPO and IA2C, along with their attention-based variants in multi-agent single-integrator environment.}
    \label{fig:comm_abla}
\end{figure}

\begin{figure}[h]
    \centering
    \includegraphics[width=0.95\linewidth]{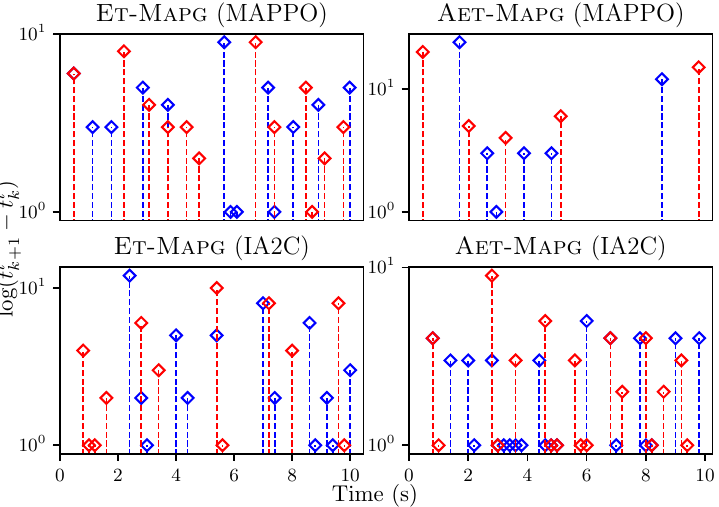}
    \caption{Inter event time of event-triggered MAPPO and IA2C, along with their attention-based variants in multi-agent single-integrator environment.}
    \label{fig:zeno_abla}
\end{figure}

To further evaluate the generality of our proposed framework, we conduct ablation studies by integrating the joint learning of event-triggered policies with control policies in IA2C~\cite{papoudakis2021benchmarking} and MAPPO~\cite{yu2022surprising}, which are two other state-of-the-art MARL algorithms. We evaluate these methods in the multi-agent single integrator environment first, owing to the fact that greater insights could be obtained from a simple environment. Since the objective in this environment is to drive the system toward the origin (the equilibrium point), event-triggered methods achieve this efficiently by reusing previously sampled actions rather than resampling at every timestep, as required in standard MARL algorithms. ~\Cref{fig:comm_abla} presents the communication frequencies of event-triggered MAPPO and IA2C, along with their attention-based counterparts. The results show that both event-triggered MAPPO and IA2C significantly reduce the number of action samples compared to their standard MARL baselines that are time-triggered. Although IA2C demonstrates weaker performance relative to MAPPO and IPPO, it still achieves over 50\% reduction in communication. \Cref{fig:zeno_abla} shows the inter-event times, verifying that triggering intervals remain strictly positive and adapt dynamically, thereby avoiding Zeno behavior. Overall, these results demonstrate that our framework is both effective and generalizable across different MARL paradigms. In particular, while MAPPO employs a centralized critic with decentralized actors and IPPO/IA2C adopts a fully independent actor–critic architecture, our event-triggered methods integrate seamlessly with these architectures while retaining their performance. 

\section{Conclusions and Future Work}\label{sec:conclusions}
In this paper, we introduced a novel framework for event-triggered MARL. We proposed \TIGER, a method that jointly learns a control action head and an event-trigger head for each agent within a single policy network, where the triggering head dynamically determines when a new action should be sampled from the action head, in contrast to prior approaches that rely on separate policies for control and triggering. Building on this, we further proposed \ATTIGER, an attention-based variant of \TIGER that incorporates self-attention as a communication mechanism during training. By enabling agents to share messages only when triggering conditions are satisfied selectively, \ATTIGER achieves efficient coordination while maintaining a sparse communication graph. Through extensive experiments across both control and standard MARL benchmarks, we demonstrated that \TIGER and \ATTIGER achieve performance comparable to MARL baselines while reducing communication and computation costs by up to 50\%. Finally, we show that our proposed methods are general and can be seamlessly integrated with any multi-agent policy gradient methods, including IPPO, MAPPO, and IA2C.

Our work also has some limitations. First, our current framework is restricted to discrete action spaces. Second, communication in \ATTIGER is assumed to occur over a complete undirected graph, where all agents share messages bidirectionally. Third, our framework is currently suitable only for policy gradient MARL methods. As future work, we plan to address these limitations by extending our methods to continuous action spaces and more complex nonlinear systems, developing techniques for communication over dynamic graphs, and extending our framework to value-based MARL approaches. 

\bibliographystyle{IEEEtran}
\bibliography{references}

\end{document}